\documentclass[10pt,a4paper,aps]{revtex4}
\usepackage{setspace}
\usepackage{parskip}
\usepackage{graphicx}
\usepackage{amssymb}
\usepackage{amsmath}
\usepackage{graphics}
\usepackage{dcolumn}
\usepackage{bm}
\usepackage{color}
\usepackage{epstopdf}
\usepackage{lipsum}
\usepackage{epstopdf}
\usepackage{hyperref} 

\begin{document}
\newcommand{\blue}[1]{\textcolor{blue}{#1}}
\newcommand{\new}{\blue}

\linespread{1.25}

	
	%
	%
	
	\title{The Effects of the Tsallis Entropy in the Proton Internal Pressure}

	\author{S. D.  Campos\footnote{email: sergiodc@ufscar.br} $~$and A. M. Amarante}

	\address{Applied Mathematics Laboratory-CCTS/DFQM, \\ Federal University of S\~ao Carlos, Sorocaba, S\~ao Paulo CEP 18052780, Brazil}
	
	
	
	
	
	\begin{abstract}
		In this paper, one discusses the effects of the Tsallis entropy on the radial pressure distribution in the proton. Using a damped confinement potential the pressure distribution is obtained from the Tsallis entropy approach, being the entropic-index $\omega$ connected with the proton temperature concerning some transition temperature. Then, the approach allows the study of the proton thermal evolution up to the Quark-Gluon Plasma regime. The von Laue stability condition, arising from the pressure distribution results in positive and negative energy regions. An analogy between the results for the radial pressure distribution and the proton-proton and the antiproton-proton total cross section is performed. The negative energy region is identified with the odderon exchange while the positive represents the pomeron exchange dominance above some transition energy $\sqrt{s_c}$. The hollowness effect is also discussed in terms of the results obtained and the proposed analogy.
		
	\end{abstract}

\maketitle

\section{Introduction}\label{sec:intro}

A central question in high energy physics is the knowledge of the topological arrangement of the quarks and gluons' inside the proton since it interferes directly with some mechanical properties, for example, pressure and density. It should be noted that radial pressure distribution, $p(r)$, in the the proton was recently measured, revealing a numerical value higher than those found in the neutron star core \cite{burkert_nature_vol_557_396_2018}. That experimental result shows two distinct pressure regions in the proton: a positive region from 0 fm up to $\sim 0.6$ fm, while the negative one resides from $\sim 0.6$ fm up to the proton edge. 

A study of the radial pressure distribution in the proton using a damped confinement potential was performed, revealing that kind of potential may be used to explain the experimental result \cite{S.D.Campos.Int.J.Mod.Phys.A34.1950057.2019}. This kind of potential can provide a parameter, the zero of the potential at $r_0$, able to give a dynamical scale for the interaction between the antiquark-quark ($\bar{q}q$) pair. If one considers a distance greater than $r_0$, then the pair is coupled as a color singlet. Otherwise, to a distance smaller than $r_0$, the quarks and antiquarks behave as a free gas. Thus, two dynamical regimes may coexist inside the proton: a strongly interacting gas, possibly obeying the Tsallis non-extensive statistic \cite{S.D.Campos.C.V.Moraes.V.A.Okorokov.Phys.Scrip.2020}, and a weakly interacting color gas, obeying the Boltzmann-Gibbs extensive statistic.

In the present work, we calculate the radial pressure distribution in the proton by taking into account the possible effects caused by the Tsallis entropy (TE). Instead of the dumped confinement potential used in Ref. \cite{S.D.Campos.Int.J.Mod.Phys.A34.1950057.2019}, one uses here a slightly modified version to describe the interaction between the $\bar{q}q$-pair. Of course, there is a myriad of confinement potentials able to furnish some information about this interaction \cite{e_eichten_Phys_Rev_Lett_34_369_1975,e_eichten_Phys.Rev.D17.3090.1978,e_eichten_Phys.Rev.D21.203.1980,a.martin.phys.lett. b93.338.1980,a.martin.phys.lett.b100.511.1988,a.martin.phys.lett.b21.561.1980,x.song.z.phys.c34.223.1987,d.b.lichtenberg.z.phys.c41.615.1989} each model being useful in some ground. 

The main role in the Tsallis statistics is played by the entropic-index $\omega$, being mathematically responsible by the measurement of the non-extensivity of the system. The non-extensivity may be interpreted as a manifestation of the strong correlation between pairs of constituents in the system. The physical definition of the entropic-index depends on the approach, and several standard results in physics can be generalized for arbitrary $\omega$ as, for example, the Langevin and Fokker-Planck equations \cite{D.A.Stariolo.Phys.Lett.A185.262.1994}, the black-body radiation Planck law \cite{C.Tsallis.F.C.Sa.Barreto.E.D.Loh.Phys.Rev.E52.1447.1995}, the Larmor precession \cite{A.R.Plastino.A.Plastino.Physica.A202.438.1994}, the localized-spin ideal paramagnetic \cite{F.D.Nobre.C.Tsallis.Physica.A213.337.1995Erratum.216.369.1995}, the study of the power-law distributions \cite{T.S.Biro.A.Jakovac.Phys.Rev.Lett.94.132302.2005,G.Wilk.Z.Wlodarczyk.AIP.Conf.Proc.1558.893.2013}, and the study of non-extensive models for heavy-ion collisions \cite{K.M.Shen.T.S.Biro.E.K.Wang.Physica.A.492.2353.2018}. It is important to stress that several entropies can be reduced to the TE \cite{beck_0902.1235v2}. 


The entropic-index $\omega$ is identified here with the temperature in the proton, which allows studying its temperature-dependent evolution up to the Hagedorn temperature \cite{R.Hagedorn.NuovoCimento.3.147.1965} near the Quark-Gluon Plasma (QGP) \cite{Y.Burnier.O.Kaczmarek.A. Rothkopf.Phys.Rev.Lett.114.082001.2015,S.Bosanyi.arXiv.1011.4230v1.2010,Y.Aok.Nature.443.675.2006}. 

As well-known, the von Laue stability condition \cite{M.von.Laue.Annalen.der.Physik.340.(8).524.1911}, roughly speaking, ensures that for non-interacting particles, the sum of positive and negative energies observed in the radial pressure distribution should cancel each other. That is a necessary condition ensuring the (mechanical) stability of the system. However, neglecting the non-interacting particle assumption, then that sum is not null \cite{M.von.Laue.Annalen.der.Physik.340.(8).524.1911}. A matter of interest is to know, at least, the sign of the von Laue condition and what it implies to the proton temperature-dependent behavior. 

On the other hand, the question about the stability condition can also be addressed more than a half-century ago in the work of Pagels \cite{H.Pagels.Phys.Rev.144.1250.1966} studying the energy-momentum tensor. In recent years, the interest was renewed and the mechanical role of this condition is studied \cite{M.V.Polyakov.C.Weiss.Phys.Rev.D60.114017.1999,K.Goeke.M.V.Polyakov.M.Vanderhaeghen.Prog.Part.Nucl.Phys.47.401.2001,O.V.Teryaev.Phys.Lett.B510.125.2001}.

The effects caused by the TE in the radial pressure distribution are analyzed through an analogy between these effects and the energy-dependence of the proton-proton ($pp$) and antiproton-proton ($\bar{p}p$) total cross section. This picture may enlighten the odderon and the pomeron roles in the high energy $pp$ and $\bar{p}p$ elastic scattering.

This paper is organized as follows. In Section \ref{sec:tbg}, one presents some basic aspects of Boltzmann-Gibbs and Tsallis entropies. In Section \ref{sec:mod}, one introduces the model used to obtain the radial pressure distribution in the proton. The stability condition is discussed according to the behavior of the $E$-term, corresponding to an energy term, in Section \ref{sec:dterm}. An analogy between the thermal evolution of the proton and the $pp$ and $\bar{p}p$ total cross section is proposed.
The final section \ref{sec:critical} deals with our remarks.

\section{The Tsallis and the Boltzmann-Gibbs Entropies: a short review}\label{sec:tbg}

The main feature of an extensive system is due to a simple property: if one divides a given system into $n$ disjoint and non-interacting cells, or at least very weakly interacting cells, then the total entropy is the sum of each cell's entropy as well as the total energy of the system. Regarding a system under thermal equilibrium, this property represents a system extensivity reduction to the additivity property of the statistical mechanics \cite{gryftopoulos_book}.

In mathematical terms, let $S$ be the total entropy of an extensive system written as
\begin{eqnarray}\label{eq:extensive}
S=S(1+2+...+W)=\sum_{i=1}^W S_i, 
\end{eqnarray}

\noindent where $W\in \mathbb{N}$ is the number of possible microscopic configurations and $S_i$ is the entropy of the $i$-th cell. The definition (\ref{eq:extensive}) is applicable for the case of the Boltzmann-Gibbs (BG) statistics, which is valid for free gas, for example. In statistical mechanics, the BG entropy is defined as the number of all possible micro-states of a given system
\begin{eqnarray}\label{eq:boltzmann}
S_B=k_B\ln W,
\end{eqnarray}

\noindent where $k_B$ is the Boltzmann constant. In a more precise explanation, one should point out that the \textit{pure} Gibbs entropy is defined over a statistical ensemble, i.e. by the distribution of all possible micro-states, resulting in slightly different entropy form. As can be easily seen from (\ref{eq:boltzmann}), the BG entropy is additive. Of course, every possible configuration is weighted by its probability of occurrence $p_i$, with the constraint
\begin{eqnarray}\label{eq:prob}
\sum_{i=1}^W p_i=1.
\end{eqnarray}

On the other hand, the non-extensive of a given system is featured by the impossibility of writing the total entropy as the sum of the entropies associated with each part of the system. A simple system composed of two independents cells, $a$ and $b$, presents a total entropy given by \cite{C.Tsallis.J.Stat.Phys.52.479.1988}
\begin{eqnarray}\label{eq:nonextensive}
S(a+b)=S(a)+S(b)+(1-\omega)S(a)S(b).
\end{eqnarray}

The parameter $\omega$ is the so-called entropic-index, representing the non-extensivity measurement of the system \cite{C.Tsallis.J.Stat.Phys.52.479.1988}. If $\omega\rightarrow 1$, then the Tsallis statistic is reduced to the BG statistic. The cases $\omega>1$ and $\omega<1$ are known as super-extensive and sub-extensive, respectively. The above definition allows the construction of the so-called Tsallis statistic \cite{C.Tsallis.J.Stat.Phys.52.479.1988}, whose leading role is to deal with non-extensive thermodynamic systems, multi-fractals, and systems where the BG statistic fails \cite{C.Beck.E.G.D.Cohen.Physica.A.322.267.2003}. The Tsallis statistic may depend on several physical constraints as, for example, the long-range interaction among the cells. The continuous form of the TE is written as \cite{S.D.Campos.C.V.Moraes.V.A.Okorokov.Phys.Scrip.2020}
\begin{eqnarray}\label{eq:tsallisentropy}
S_T(x,\omega)=\frac{k_B}{\omega-1}\left(1-\int_{-\infty}^\infty p(x)^{\omega}dx\right),
\end{eqnarray}

\noindent and the probability can also depends on the entropic-index, $p(x)=p(x,\omega)$ for fixed $\omega$. The TE can also be reduced to the BG entropy in the limit $\omega\rightarrow 1$.

The choice of the ensemble is a fundamental starting point for the problem. One uses here the canonical ensemble, where the system and the reservoir are in thermal equilibrium, and there is no particle exchange between the reservoir and the system. In this ensemble, the usual probability of a given micro-state can be written
\begin{eqnarray}\label{eq:bgprob}
p_i=e^{-\frac{F- \epsilon_i}{k_B T}},
\end{eqnarray}

\noindent where $\epsilon_i$ is the internal energy in the $i$-th cell, $T$ is the temperature, and $F$ is the constant Helmholtz free energy. The exponential function written above is the usual one, defined as
\begin{eqnarray}\label{eq:exp}
e^x
=\lim_{n\rightarrow\infty}\left(1+\frac{x}{n}\right)^n.
\end{eqnarray}

The probability (\ref{eq:bgprob}) should be rewritten to take into account the modifications needed by the introduction of the entropic-index. In the Tsallis statistic, the canonical ensemble can be written using the probability $p(x,\omega)$ as \cite{A.S.Parvan.Eur.Phys.J.A51.108.2015} (hereafter $k_B=1, c=1,\hslash=1$)
\begin{eqnarray}\label{eq:probgrandcanon}
p(x,\omega)=\left[1+\frac{1-\omega}{ T}(F-E(x))\right]^{\frac{1}{\omega-1}}.
\end{eqnarray}

The term in the r.h.s of (\ref{eq:probgrandcanon}) is the so-called $\omega$-exponential
\begin{eqnarray}\label{eq:qexp}
e_\omega^x=[1+(1-\omega)x]^{\frac{1}{1-\omega}},
\end{eqnarray}

\noindent and for the limiting case $w\rightarrow 1$, one has
\begin{eqnarray}\label{eq:qexplimit}
\lim_{\omega\rightarrow 1}e_\omega^x=\lim_{\omega\rightarrow 1}[1+(1-\omega)x]^{\frac{1}{1-\omega}}=e^x,
\end{eqnarray}

\noindent and, therefore, the desirable case $\omega\rightarrow 1$ brings us the BG entropy as a particular case of the TE. 

As well-known, the TE is one of a wide class of possible ways to compute the entropy of a given system \cite{C.E.Shannon.Bell.S.Tech.J.27.379.1948,A.Renyi.Proc.4th.Berkeley.Symp.on.Mathematics.Statistics.and.Probability.p.547.1960,T.S.Biro.G.G.Barnafoldi.P.Van.Physica.A.417.215.2015}. The model introduced in the next section takes into account the strong interaction between $\bar{q}q$-pairs in the proton (antiproton), which turns the proton (antiproton) a non-extensive system. Thus, the TE can be used, bringing information on how the radial pressure is affected by the entropy.

\section{The Model}\label{sec:mod}

For the sake of simplicity, one considers the proton as composed only by $\bar{q}q$-pairs, depending on the relative spatial distance $r$. Then, this distance can vary from 0 fm up to $2r_p$, where $r_p$ is the experimental proton radius. Hereafter, one assumes a constant proton radius, $r_p=0.5$ fm. That is, of course, a simplification of the problem since the radius, probably, is a temperature-dependent function. 

\subsection{The Damped Confinement Potential}

The QCD is not a trivial theory since its very beginning \cite{M.Gell-Mann.book.1999}. Even the definition of its vacuum from the so-called QCD sum rules is a real challenge \cite{M.A.Shifman.M.Kislinger.V.I.Sakharov.Nucl.Phys.B147.448.1978}. However, the asymptotic freedom of QCD \cite{D.J.Gross.F.Wilczek.Phys.Rev.Lett.30.1343.1973,H.D.Politzer.Phys.Rev.Lett.30.1346.1973} seems quite simple when applied to the hard processes. Considering short distances the effective coupling constant of QCD becomes small, and the interaction can be treated by using perturbative techniques. Long distances, however, prevents the use of perturbative techniques. At this point, the introduction of the potential approach to understand the confinement/non-confinement regime, despite its simplicity, may allow at least a first glance on general features of the pair interaction, and what effects one can expect from some limiting situations.

In the potential approach, depending on the distance $r$ between the $\bar{q}q$-pairs, they can be in the confinement or the non-confinement regime. Then, considering some scale $r_0$, for $r<r_0$ implies the pairs are in the non-confinement regime, behaving as a free gas; for $r>r_0$, they are in the confinement regime, behaving as a color singlet. The scale $r_0$ is defined here from the confinement potential under study. 

The dumped Cornell confinement potential is written here as \cite{S.D.Campos.Int.J.Mod.Phys.A34.1950057.2019}
\begin{eqnarray}\label{eq:confin}
V(r)=n\left(-2\sigma\sqrt{\frac{\alpha_s(\mu^2)}{3\sigma}}+\sigma r\right)e^{\frac{\lambda_D}{r-2r_p}},
\end{eqnarray}

\noindent where $n$ is the finite number of pairs at distance $r$ (hereafter $n=1$), $\sigma$  is the string tension \cite{PRD-90-074017-2014}, $\lambda_D$ is the Debye length, and it is about $0.15$ fm in the QGP regime \cite{Y.Burnier.O.Kaczmarek.A. Rothkopf.Phys.Rev.Lett.114.082001.2015,S.Bosanyi.arXiv.1011.4230v1.2010,Y.Aok.Nature.443.675.2006}. The $\alpha_s(\mu^2)$ is the running coupling constant responsible by the strong interaction at a specific energy scale $\mu$, written in the one-loop approximation \cite{PDG-PhysRev-D98-030001-2018}, as
\begin{eqnarray}\label{eq:rcc}
\alpha_s(\mu^2)=\frac{1}{4\pi\beta \ln\left(\frac{\mu^2}{\Lambda_{QCD}^2}\right)}, 
\end{eqnarray}

\noindent with $\beta_{0} = (33-2n_{f})/12\pi^2$ being the $1$-loop $\beta$-function. The number of active quark flavors at $\mu$ is given by $n_{f}$, and are considered light $m_{q} \ll \mu$. The $m_{q}$ is the quark mass and the subscript $q$ its flavor: $n_f=6$ for $\mu\geq m_t$, $n_f=5$ for $m_b\leq \mu \leq m_t$, $n_f=4$ for $m_c\leq \mu \leq m_b$ and $n_f=3$ for $\mu\leq m_c$ \cite{buras_book}. The $\Lambda_{\scriptsize{\mbox{QCD}}}$-parameter depends on the renormalization scheme and on the flavor number $n_{f}$ \cite{PDG-PhysRev-D98-030001-2018}. The scale is given by transition value $r_0$, written as
\begin{eqnarray}
r_0=2\sqrt{\frac{\sigma}{3\alpha_s(\mu^2)}},
\end{eqnarray}

\noindent and it is just defined as the zero of potential (\ref{eq:confin}). The definition of the energy scale $\mu$ depends on the system properties. For example, it is possible to identify $\mu^2=Q^2$, where $Q^2$ the transferred momentum in the gluon frame \cite{S.D.Campos.Int.J.Mod.Phys.A34.1950057.2019}. The non-relativistic regime demands large $Q^2$ (small distances). Furthermore, the existence of the confinement phase is related to the growth of $\alpha_s(Q^2)$ at small $Q^2$ (large distances). On the other hand, in a recent paper \cite{S.D.Campos.C.V.Moraes.V.A.Okorokov.Phys.Scrip.2020}, one introduces the entropic-index as the ratio
\begin{eqnarray}\label{eq:ratio}
\omega'=\frac{s}{s_c},
\end{eqnarray}

\noindent where $s$ is the squared energy in the center of mass system and $s_c$ is the critical value associated with the minimum of the hadronic total cross section \cite{S.D.Campos.C.V.Moraes.V.A.Okorokov.Phys.Scrip.2020}. In that picture, the factor $(s/s_c-1)^{-1}=(\omega'-1)^{-1}$ is fundamental to understand the change in the sign of the TE as the energy increases.

On the other hand, the radial pressure distribution, $p(r)$, in the proton was recently measured using the deeply virtual Compton scattering, where high-energy electrons were scattered from the proton in liquid hydrogen \cite{burkert_nature_vol_557_396_2018}. In that kind of scattering $s<\!\!<s_c$ and, then, the TE is negative. Moreover, a change in the sign of the entropy may imply changes in the sign of the radial pressure. Then, the $g'$ factor used in \cite{S.D.Campos.Int.J.Mod.Phys.A34.1950057.2019} was defined as
\begin{eqnarray}\label{eq:gfactor}
g'=\frac{1}{s/s_c-1},
\end{eqnarray}

\noindent which result in $g'=-1$ for $s<\!\!<s_c$. Therefore, this factor is responsible for takes into account the possible phase-transition shown by the $pp$ and $\bar{p}p$ total cross section experimental data, and it also gives the correct profile for the radial pressure distribution in the proton \cite{S.D.Campos.Int.J.Mod.Phys.A34.1950057.2019}.

The canonical ensemble is featured by the absence of particle exchange and by the presence of thermal equilibrium. If one considers each $\bar{q}q$-pair as possessing a specific $\omega$, then the simple definition of temperature can be a challenge \cite{M.Nauenberg.Phys.Rev.E.67.036114.2003}. For the sake of simplicity, however, one introduces only one entropic-index for all interacting pairs and depending only on the internal temperature in straight analogy with (\ref{eq:ratio})
\begin{eqnarray}\label{eq:entropicTC}
\omega=\frac{T}{T_c},
\end{eqnarray}

\noindent where $T_c$ is some transition temperature at the minimum of the total cross section. It is interesting to point out the entropic-index can be related with the heat reservoir capacity as well as with the temperature fluctuations \cite{T.S.Biro.Physica.A.392.3132.2013,G.B.Bagci.T.Oikonomou.Phys.Rev.E.88.042126.2013,T.S.Biro.G.G.Barnafoldi.P.Van.Physica.A.417.215.2015,K.M.Shen.T.S.Biro.E.K.Wang.Physica.A.492.2353.2018} of the system. In particular, the identification of $\omega$ with temperature fluctuations can be used to study multiplicity fluctuations in high-energy collisions \cite{G.Wilk.Z.Wlodarczyk.Eur.Phys.J.A.40.299.2009,G.Wilk.Z.Wlodarczyk.Eur.Phys.J.A.48.161.2012}.

Based on the $pp$ and $\bar{p}p$ total cross section behavior, it is reasonable to suppose a transition temperature below the Hagedorn temperature, which may explain the dramatic change in the total cross section behavior shown by the ISR measurement, and not predicted by any model. As well-known, the Hagedorn temperature $T_H\approx 185$ MeV is defined as the melting hadron temperature, and can also be viewed as a mass scale of the order of the pion mass, $T_H\approx m_\pi$. Then, taking into account the above discussion, the $g$-term is defined as
\begin{eqnarray}\label{eq:gterm}
g=\frac{1}{T/T_c-1}=\frac{1}{\omega -1},
\end{eqnarray}

\noindent where $T_c<T_H$. In the present case, the proton temperature is far below $T_H$, representing the situation where the $\bar{q}q$-pairs may be strongly correlated, depending on the pair formation distance $r$. Then, $\omega\neq 1$, which means a non-extensive system allowing the use of the TE. However, as the temperature increases, the hadron may suffer some phase-transition, analogous to that described in Ref. \cite{S.D.Campos.C.V.Moraes.V.A.Okorokov.Phys.Scrip.2020}. Without loss of generality,  one defines $T_c=1$ MeV and $T_H=2$ MeV, writing for a given temperature $T$, 
\begin{eqnarray}\label{eq:truque}
T=\frac{T_c}{T_c}T=T_c\omega := \Omega.
\end{eqnarray}

Note that $\omega$ is dimensionless and $\Omega$ has unit of energy.

The internal energy of the system is represented by $U(r)$ and it takes into account all kinds of energy. The precise knowledge of all forms of energy in the system is not accessible from the theoretical point of view. However, due to the strong interaction one can suppose the main energy form is given by the potential (\ref{eq:confin}), neglecting the kinetic term as well as the action due to external forces. Despite its naivety, this approach seems to agree, at qualitative level, to the thermodynamic $T$-matrix formalism \cite{Sh.F.Y.Liu.R.Rapp.arXiv:1501.07892.hep-ph.2015}. The quantum potential due to Bohm can be used to mimic the internal energy of a quantum system, giving an insight into its role in stationary states \cite{G.Dennis.M.A.de.Gosson.B.J.Hiley.Phys.Lett.A379.1224.2015,G.Dennis.M.A.de.Gosson.B.J.Hiley.Phys.Lett.A378.2363.2014}. Then, from the Bohm’s point of view, the particle is not a point-like object, possessing an internal structure with some topological geometry. As shall be seen, this extended structure is necessary to explain the hollowness effect.

\subsection{The Normalized Probability}

The probability (\ref{eq:probgrandcanon}) should be normalized by its maximum value, resulting the normalized probability $\bar{p}(r,\omega)$
\begin{eqnarray}\label{eq:probmax}
\bar{p}(r,\omega)=\frac{p(r,\omega)}{p(r_{max},\omega)}=\frac{1}{p_{max}}\left[1+\frac{1-\omega}{\Omega}(F-U(r))\right]^{\frac{1}{\omega -1}},
\end{eqnarray}   

\begin{figure}
	\centering{
		\includegraphics[scale=0.4]{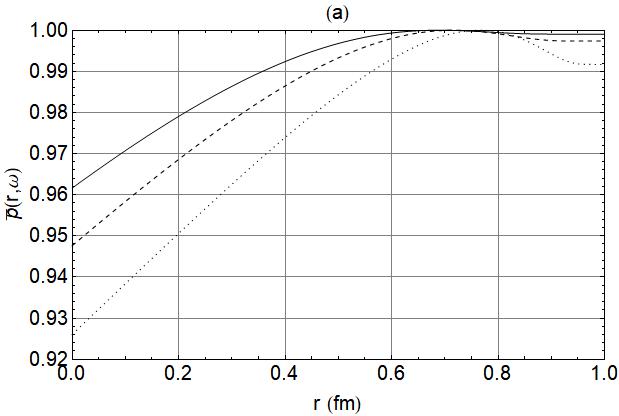}
		\includegraphics[scale=0.4]{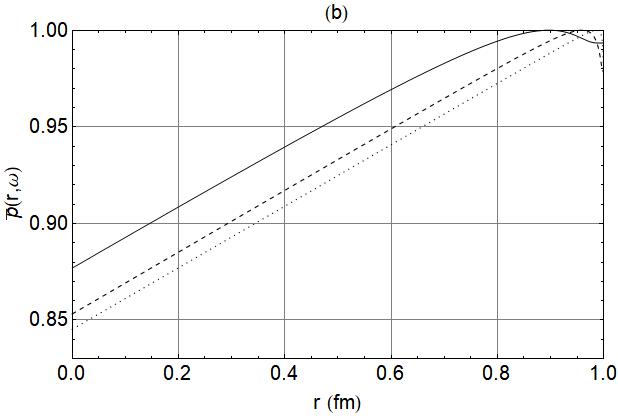}
		\caption{The normalized probability (\ref{eq:probmax}) as a function of the $\bar{q}q$-pair separation $r$. In panel (a) $Q=20$ GeV and $\omega=0.001$ for all curves: $\lambda=0.9$ fm (solid-line), $\lambda=0.6$ fm (dashed-line) and $\lambda=0.3$ fm (dotted-line). In panel (b), $Q=3$ GeV and $\omega=2$ fm for all curves: $\lambda=0.1$ fm (solid-line), $\lambda=0.01$ fm (dashed-line) and $\lambda=0.001$ fm (dotted-line). In both panels $n_f=3$, $\Lambda=0.309$ GeV, $\sqrt{\sigma}=0.405$ GeV, and $r_p=0.5$ fm.}
		\label{fig:fig_prob_norm}}
\end{figure}

\noindent where $p(r_{max},\omega)=p_{max}$ corresponds to the maximum of the probability for an internal energy $U(r)$. Neglecting the kinetic energy of the $\bar{q}q$-pair, one identifies the internal energy and the damped potential assuming $U(r)=V(r)$. Moreover, $r_{max}$ is given by
\begin{eqnarray}\label{eq:rmax}
r_{max}=\frac{12\pi r_p+3\pi\lambda-\sqrt{3 \pi}\sqrt{ 24\pi r_p\lambda+3\pi\lambda^2-4\sqrt{3\pi}\lambda\sqrt{\frac{\alpha_s(Q^2)}{\sigma}}}}{6 \pi}.
\end{eqnarray}

The normalized probability obtained above produces
\begin{eqnarray}\label{eq:probnorm}
\int_0^{2r_p}\bar{p}(r,\omega)dr \approx 1,
\end{eqnarray}

\noindent as can be seen from the results shown in Figure \ref{fig:fig_prob_norm}. The changing in $\omega$ according to $T$ introduces the need for a variation in the Debye length since $\lambda$ is important to characterize the system as relativistic or non-relativistic. Then, in Figure \ref{fig:fig_prob_norm}a, the entropic-index is assumed constant, $\omega=0.001$, which means a temperature very far from the transition temperature. In this panel, the Debye length is taken as 0.9 fm (solid-line), 0.6 fm (dashed-line) and 0.3 fm (dotted-line). 
In panel (b), one takes $\omega=2$, representing the saturation of the Hagedorn temperature. In this situation the Debye length is 0.1 fm (solid-line), 0.01 fm (dashed-line) and 0.001 fm (dotted-line). 

Therefore, the normalized probability (\ref{eq:probnorm}) is robust under extreme variations of $\omega$ and $\lambda$. Hereafter, one uses $\lambda=0.9$ fm in the non-relativistic regime and $\lambda=0.1$ fm in the relativistic regime.



\subsection{The Tsallis Entropy}

Using the normalized probability (\ref{eq:probmax}), one can write the continuous TE as
\begin{eqnarray}\label{eq:tsallis_dcp}
S(r,\omega)=\frac{1}{\omega-1}\left(1-\frac{1}{p_{max}}\int_0^{2r_p}f(r,\omega)dr\right),
\end{eqnarray}

\noindent where 
\begin{eqnarray}
f(r,\omega)=\left[1+\frac{1-\omega}{\Omega}\left[F-\left(-2\sigma\sqrt{\frac{\alpha_s(\mu^2)}{3\sigma}}+\sigma r\right)e^{\frac{\lambda_D}{r-2r_p}}\right]\right]^{\frac{\omega}{\omega-1}}.
\end{eqnarray}

\begin{figure}
	\centering{
		\includegraphics[scale=0.4]{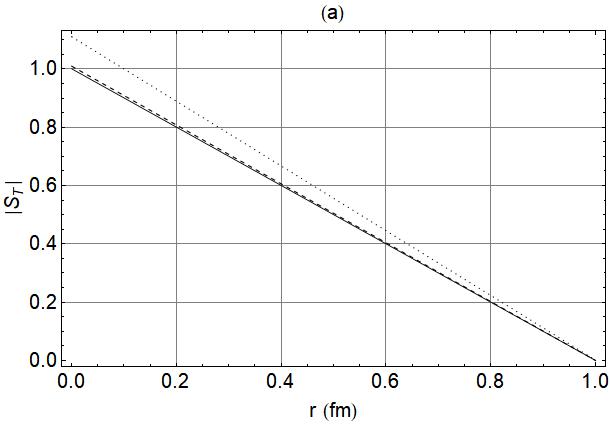}
		\includegraphics[scale=0.4]{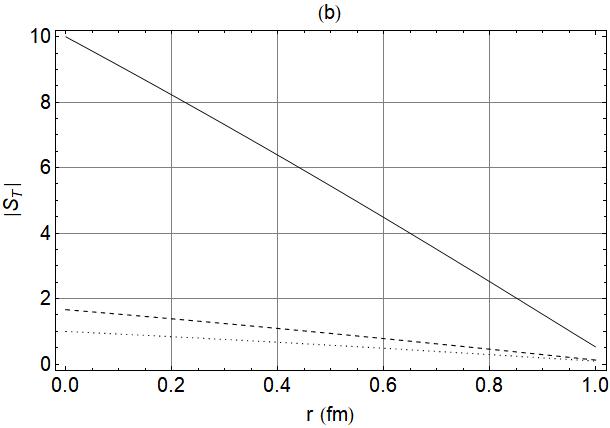}
		\caption{The absolute value of the TE (\ref{eq:tsallis_dcp}) as a function of the $\bar{q}q$-pair separation $r$. The numerical integration is performed assuming $n_f=3$, $\Lambda=0.309$ GeV, $\sqrt{\sigma}=0.405$ GeV, $r_p=0.5$ fm. In panel (a), $Q=20$ GeV and $\lambda=0.9$ fm: $\omega=0.001$ (solid-line), $\omega=0.01$ (dashed-line), $\omega=0.1$ (dotted-line). In panel (b), $Q=3$ GeV, $\lambda=0.1$ fm: $\omega=1.3$ (solid-line), $\omega=1.6$ (dashed-line), and $\omega=2.0$ (dotted-line).}
		\label{fig:fig_tsallis_dcp}}
\end{figure}

Unfortunately, the result (\ref{eq:tsallis_dcp}) cannot be analytically integrated. Figure \ref{fig:fig_tsallis_dcp}a shows the absolute value of the TE assuming $\lambda=0.9$ fm. The solid-line is for $\omega=0.001$, dashed-line is for $\omega=0.01$, and dotted-line is for $\omega=0.1$. In Figure \ref{fig:fig_tsallis_dcp}b, one uses $\lambda=0.1$ fm . The solid-line is for $\omega=1.3$, the dashed-line is for $\omega=1.6$, and dotted-line is for $\omega=2$. 

It is possible to observe that, as the pair separation distance decreases, the number of degrees of freedom rises, which is expected to occur when the system is through from the confinement to the non-confinement phase. It implies the TE grows while the pair separation distance decreases up to its minimum value. On the other hand, the TE decreases as the pair separation distance grows, caused by the diminishing of the number of degrees of freedom. 

\section{The Radial Pressure Distribution and the E-Term}\label{sec:dterm}

The pressure distribution in the system under study can be obtained from the fundamental thermodynamic relation 
\begin{eqnarray}\label{eq:fundrelation}
dU(r)=-S_TdT-PdV_{ol}(r)+\mu dn,
\end{eqnarray}

\noindent where $P=P(r,\omega)$ is the pressure distribution, $V_{ol}(r)$ is the hadron volume, and $\mu$ is the chemical potential. One adopts $\mu=0$ and $V_{ol}(r)\approx r^3$. It is important to stress that $P(r,\omega)$ is finite at the origin, however, $P(0,\omega)\neq 0$. Then, one writes the following pressure only to ensure a null pressure at $r=0$
\begin{eqnarray}\label{eq:pres_alt}
\tilde{P}(r,\omega)=\frac{r}{2r_p}P(r,\omega).
\end{eqnarray}

Note this procedure only affects the radial pressure at the origin since at the proton edge $r=2r_p$, then $\tilde{P}(r,\omega)=P(r,\omega)$. Thus, there is no change in the physics of the problem. As stated in \cite{S.D.Campos.Int.J.Mod.Phys.A34.1950057.2019}, the $g$ factor should be taken into account to give the correct internal pressure distribution sign. Therefore, following the above prescriptions, it is straightforward to obtain from (\ref{eq:tsallis_dcp}), (\ref{eq:fundrelation}) and (\ref{eq:pres_alt}) the radial pressure distribution $\tilde{P}(r,\omega)$ (for $T$ constant)
\begin{eqnarray}\label{eq:pressure}
r^2\tilde{P}(r,\omega)=-\frac{r}{2r_p}\left(g\frac{\partial V(r)}{\partial r}+\Omega \frac{\partial S_T}{\partial r}\right).
\end{eqnarray}

Note that $ g $ factor given by (\ref{eq:entropicTC}) is encompassed in the second term in the parentheses of (\ref{eq:pressure}).

\begin{figure}
	\centering{
		\includegraphics[scale=0.4]{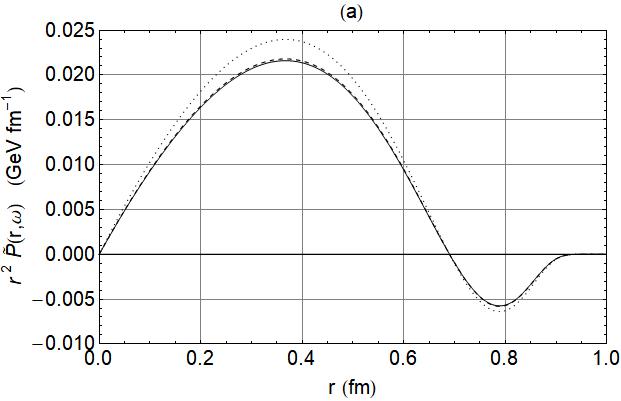}
		\includegraphics[scale=0.375]{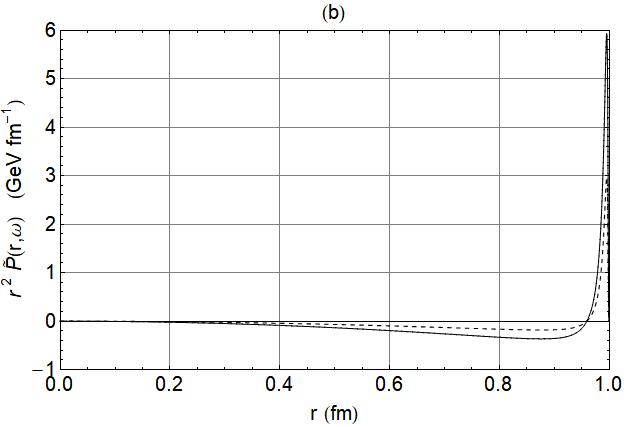}
		\caption{Radial pressure distribution in the proton using (\ref{eq:pressure}) and neglecting the TE. The panels (a) and (b) show a pattern quite similar to those achieved in \cite{S.D.Campos.Int.J.Mod.Phys.A34.1950057.2019}. In panel (a), one uses $Q=20$ GeV and $\lambda=0.9$ fm: $\omega=0.001$ (solid-line), $\omega=0.01$ (dashed-line), and $\omega=0.1$ (dotted-line). In panel (b), one uses $Q=3$ GeV and $\lambda=0.1$ fm: $\omega=1.3$ (solid-line), $\omega=1.6$ (dashed-line), and $\omega=2.0$ (dotted-line). For both panels, one uses $n_f=3$, $\Lambda=0.309$ GeV, $\sqrt{\sigma}=0.405$ GeV, and $r_p=0.5$.}
		\label{fig:fig_press_1}}
\end{figure}

A necessary condition for the proton stability is given by the well-known von Laue condition \cite{M.von.Laue.Annalen.der.Physik.340.(8).524.1911} 
\begin{eqnarray}\label{eq:vonLaue_1}
\int_0^{\infty}r^2\tilde{P}(r)dr=0,
\end{eqnarray}

\noindent which is obtained by considering a system composed of non-interacting particles. That kind of system is subject to the BG statistic. An alternative assumption, given also by von Laue, is to assume that the surrounding exerts an equal pressure from all sides \cite{M.von.Laue.Annalen.der.Physik.340.(8).524.1911}, allowing that particles can interact with each other. Then, one writes (\ref{eq:vonLaue_1}) in a more general way
\begin{eqnarray}\label{eq:vonLaue_2}
\int_0^{\infty}r^2\tilde{P}(r,\omega)dr= E,
\end{eqnarray}

\noindent where $E$ has unit of energy \cite{M.von.Laue.Annalen.der.Physik.340.(8).524.1911}. The system under study assumes a finite size to the proton as well as a finite range to the strong interaction, imposing a severe constraint on (\ref{eq:vonLaue_1}) and (\ref{eq:vonLaue_2}). Then, the stability condition (\ref{eq:vonLaue_2}) only holds as approximation (for non-interacting particles)
\begin{eqnarray}\label{eq:vonLaue_3}
\int_0^{r_p}r^2\tilde{P}(r)dr\approx 0,
\end{eqnarray}

\noindent being null only under some conditions. In particular, it holds if the damped confinement potential is symmetric regarding $r_0$ in the sense that negative and positive contributions are equal in modulus. As can be easily seen from (\ref{eq:confin}), the damped confinement potential adopted here is not symmetric regarding $r_0$. Then, the sum of positive and negative regions shown in Figure \ref{fig:fig_press_1} is not exactly zero. 

Figure \ref{fig:fig_press_1}a and \ref{fig:fig_press_1}b shows the radial pressure distribution in the proton coming only from the damped potential, i.e. neglecting the TE contribution. In panel (a), one has $T<T_c$ for all curves, and in panel (b), $T>T_c$. Note that a phase transition takes place at $T_c$, which implies the change in the sign of the radial pressure distribution. 

\begin{figure}
	\centering{
		\includegraphics[scale=0.41]{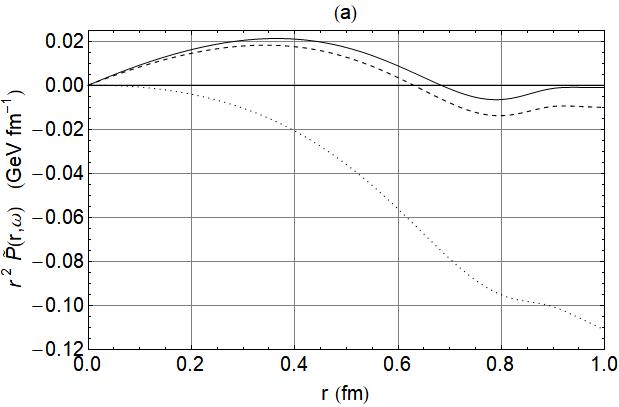}
		\includegraphics[scale=0.4]{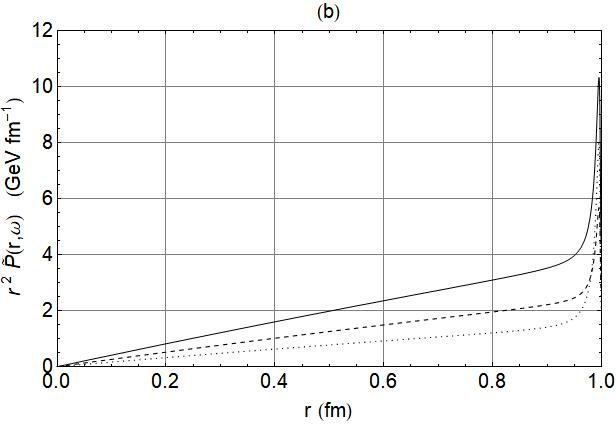}
		\caption{Radial pressure distribution in the proton using (\ref{eq:pressure}) and taking into account the TE contribution. For the panel (a), one uses $Q=20$ GeV fm and $\lambda=0.9$ fm: $\omega=0.001$ (solid-line), $\omega=0.01$ (dashed-line), and $\omega=0.1$ (dotted-line). In panel (b), one uses $Q=3$ GeV and $\lambda=0.1$ fm: $\omega=1.3$ (solid-line), $\omega=1.6$ (dashed-line), and $\omega=2.0$ (dotted-line).	For both panels, one uses $n_f=3$, $\Lambda=0.309$ GeV, $\sqrt{\sigma}=0.405$ GeV, and $r_p=0.5$.}
		\label{fig:fig_press_2}}
\end{figure}

In the present case under study, the von Laue condition holds as an approximation 
\begin{eqnarray}\label{eq:vonLaue_4}
\int_0^{r_p}r^2\tilde{P}(r)dr\approx E,
\end{eqnarray}

\noindent and $E=E(\omega)$. The sign of $E$ depends on the result of the sum of positive and negative regions. In Figure \ref{fig:fig_press_2}a and \ref{fig:fig_press_2}b, one takes into account the TE contribution to the radial pressure distribution in the proton. The sharp peak shown in the panel \ref{fig:fig_press_2}b near the proton edge is caused by the maximum of the probability (\ref{eq:probmax}). The results are shown in Figure \ref{fig:fig_press_2}, and can be interpreted assuming, as extrapolation, that proton radius can vary according to the energy sign in the von Laue stability condition.

Considering $T<T_c$, the TE implies the rise of the negative energy region (in modulus), leading to the vanishing of the positive region for some $\omega_0<\omega$. Note that greater is the entropic-index $\omega_0<\omega\rightarrow 1$, greater is the negative energy region (in modulus). One may consider here that negative energy can be defined analogously to the gravitational energy, i.e. as attractive. Then, the dominance of the negative energy may lead, eventually, to a decreasing of the proton radius up to some limiting value, achieved when $\omega\rightarrow 1$. This minimum value to the proton radius may be a consequence of the Pauli blocking effect, i.e. the fact fact that Pauli principle prevents the multiple occupation of quark states. 

When $T_c<T \Rightarrow 1<\omega$, then the negative region vanishes and only the positive region survives. Thus, the TE may lead, eventually, to the rise of the positive energy region for $1<\omega$. Now, the positive energy is viewed as repulsive, possibly implying the growth of the proton radius. The process should cease when $T_c\lesssim T\rightarrow T_H$, which means the $\bar{q}q$-pairs has achieved the QGP regime.

The region near the phase transition can be studied according to the approach performed in high-energy collisions \cite{T.S.Biro.A.Jakovac.Phys.Rev.Lett.94.132302.2005,G.Wilk.Z.Wlodarczyk.Eur.Phys.J.A.48.161.2012,G.B.Bagci.T.Oikonomou.Phys.Rev.E.88.042126.2013,T.S.Biro.G.G.Barnafoldi.P.Van.Physica.A.417.215.2015,T.S.Biro.Physica.A.392.3132.2013}. Thus, one can assume temperature fluctuations near $T=T_c$ redefining the entropic-index by using \cite{G.Wilk.Z.Wlodarczyk.Eur.Phys.J.A.48.161.2012} 
\begin{eqnarray}\label{eq:wilk_1}
	\omega=1+\frac{\mathrm{Var}(T)}{\langle T\rangle^2},
	\end{eqnarray}

\noindent where $\mathrm{Var}(T)$ is the variance of $T$. This procedure may be interesting to investigate the dominance of negative/positive energy near the transition temperature and will be developed elsewhere.

\subsection{Total Cross Section and Thermal Evolution: An Analogy}

The above results can be reanalyzed through an analogy with the $pp$ and $\bar{p}p$ total cross section, $\sigma_{tot}^{pp}(s)$ and $\sigma_{tot}^{\bar{p}p}(s)$, respectively. One considers these total cross sections are governed by the odderon/pomeron exchange above the Coulomb and Nuclear interference region.

Considering $s<s_c$, the total cross section is a decreasing function of $s$, and at $s=s_c$ the total cross section achieves its minimum experimental value as can be viewed in the experimental data \cite{PDG-PhysRev-D98-030001-2018}. It is important to stress that $s_c$ is not necessarily the same for $\sigma_{tot}^{pp}(s)$ and $\sigma_{tot}^{\bar{p}p}(s)$. In this energy regime, it is assumed the odderon is the leading particle exchange \cite{S.D.CamposPhys.Scr.95.065302.2020,S.D.Campos.arxiv.2020}. As well-known, the odderon has $C=-1$ parity, which explains the different patterns observed in the $\sigma_{tot}^{pp}(s)$ and $\sigma_{tot}^{\bar{p}p}(s)$ experimental data in this energy range \cite{S.D.CamposPhys.Scr.95.065302.2020}.

These patterns, from the TE point of view, may be attributed to the different topological arrangement of the $\bar{q}q$-pairs inside the proton and antiproton. One may attribute a soft correlation to the $\bar{q}q$-pairs inside the proton implying that when ceases the Coulomb-Nuclear interference, $\omega\approx 1$. Thus, the correlation term in (\ref{eq:nonextensive}) has a small contribution to the TE inside the proton. Then, the entropy generated by the $\bar{q}q$-pair interaction in the $pp$ picture may be given in good approximation by the BG entropy.

On the other hand, if one assumes the correlation inside the antiproton is strong, then $\omega<\!<\!1$. Therefore, the correlation term in (\ref{eq:nonextensive}) cannot be disregarded for the $\bar{p}p$ case. The differences between these topological arrangements tend to disappear with the increasing energy as can be observed in the experimental data set for $pp$ and $\bar{p}p$ \cite{PDG-PhysRev-D98-030001-2018}. 

These assumptions may explain the $pp$ total cross section slow decreasing toward the minimum shown by the experimental data. Indeed, the experimental data for $pp$ total cross section is almost flat compared to the $\bar{p}p$ in this energy range \cite{PDG-PhysRev-D98-030001-2018}.  

On the other hand, considering $s_c< s$, the pomeron exchange starts to be the main particle exchange. The pomeron has $C=+1$ parity, turning particles and antiparticles the same entity from its point of view. Then, the $pp$ and $\bar{p}p$ total cross section tends to has the same behavior, which means the same topological arrangement as $s_c<s\rightarrow\infty$ (Pomeranchuk theorem \cite{I.Ia.Pomeranchuk.Sov.Phys.JETP.7.499.1958}).

Considering the analogy performed above, entropic-index $\omega$ may be viewed as a measure of the odderon/pomeron dominance, according to the collision energy range. The odderon exchange may be viewed as acting to diminish the $pp$ and $\bar{p}p$ total cross section. From the thermodynamic point of view, the odderon exchange is represented by the dominance of the negative energy in the stability condition. Then, the dominance of the odderon exchange means the rise of the negative energy region shown in Figure \ref{fig:fig_press_2}a. 

When the positive energy region turns to be dominant for $1<\omega$, then the pomeron becomes the leading particle exchange. The pomeron represents the positive energy region in Figure \ref{fig:fig_press_2}b. Therefore, the pomeron exchange implies the growth of the total cross section as $s_c<s\rightarrow\infty$.

Another important question is the existence or not of the hollowness effect \cite{I.M.Dremin.Phys.Usp.58.61.2015,A.Alkin.E.Martynov.O.Kovalenko.S.M.Troshin.Phys.Rev.D89.091501.2014,S.M.Troshin.N.E.Tyurin.Mod.Phys.Lett.A31.1650079.2016,S.M.Troshin.N.E.Tyurin.Eur.Phys.J.A53.57.2017,I.M.Dremin.Phys.Usp.60.333.2017,V.V.Anisovich.V.A.Nikonov.J.Nyiri.Phys.Rev.D90.074005.2014,W.Broniowski.E.RuizArriola.Acta.Phys.Polon.B.Proc.Suppl.10.1203.2017,E.RuizArriola.W.Broniowski.Phys.Rev.D95.074030.2017,S.D.Campos.V.A.Okorokov.2018,T.Csorgo.R.Pasechnik.A.Ster.Eur.Phys.J.C.80.126.2020}. If one disregard the TE, then there is a non-vanishing negative region for $1<\omega$ in the damped potential approach \cite{S.D.Campos.Int.J.Mod.Phys.A34.1950057.2019}, which may represent the hollowness effect, which has thermodynamic origin \cite{S.D.Campos.V.A.Okorokov.2018}. In the black disc picture, the disc becomes blacker as the collision energy grows, diminishing the negative energy region in the stability condition. However, there is a non-null negative region, turning the black disc less black than the classical picture.

On the other hand, considering the presence of the TE, then the hollowness effect vanishes for $1<\omega$. Thus, the hollowness effect can exist only by the emergence of some blocking mechanism preventing the vanishing of the negative energy region. For example, small fluctuations near the proton core may lead to the formation of negative energy regions. Far from the proton core, positive energy is strong enough to prevent the formation of such structures. Assuming that the negative energy region corresponds to the odderon exchange, then such small regions near the core may represent the odderon exchange, which may contribute to tame the rise of the total cross section. Therefore, the presence of the hollowness effect may be viewed as evidence of the (small) odderon exchange at very high energies.

The analogy between the thermal evolution of the proton and the energy-dependent behavior of the $pp$ and $\bar{p}p$ total cross section, on the other hand, may help to understand the role played by the odderon/pomeron exchange. However, the potential approach does not take into account the subtleties of the odderon/pomeron dominance near the transition energy $s_c$, where the total cross section achieves its minimum experimental value. For example, energy fluctuations near the transition temperature can occur,  being possibly important to explain the mixed region, where the odderon and the pomeron exchanges may be relevant to describe the total cross section. In this way, the approach performed in \cite{T.S.Biro.A.Jakovac.Phys.Rev.Lett.94.132302.2005,G.Wilk.Z.Wlodarczyk.Eur.Phys.J.A.48.161.2012,G.B.Bagci.T.Oikonomou.Phys.Rev.E.88.042126.2013,T.S.Biro.G.G.Barnafoldi.P.Van.Physica.A.417.215.2015,T.S.Biro.Physica.A.392.3132.2013} can be relevant to study the temperature fluctuations near $T_c$.

\section{Critical Remarks}\label{sec:critical}

This paper presents a model based on a damped confinement potential able to explain the radial pressure distribution in the proton. It is performed an analysis considering the TE as being produced only by the $\bar{q}q$-pairs inside the proton.

One uses a temperature-dependent entropic-index. A transition temperature $T_c$ is assumed to occur at the minimum of the total cross section, below the Hagedorn temperature $T_H$. This assumption is based on the behavior of $\sigma_{tot}^{pp}(s)$ and $\sigma_{tot}^{\bar{p}p}(s)$ experimental data, which exhibits a minimum at some $s_c$ (not necessarily unique). The adoption of a temperature-dependent entropic-index is not new in high-energy physics \cite{T.S.Biro.Physica.A.392.3132.2013,T.S.Biro.G.G.Barnafoldi.P.Van.Physica.A.417.215.2015}.

The radial pressure distribution in the proton can be analyzed through the von Laue stability condition \cite{M.von.Laue.Annalen.der.Physik.340.(8).524.1911}, where the energy $E$ has a crucial role. This term represents the dominance of the positive or negative sign in the stability condition. 

The negative energy is attractive and dominant for $\omega_0<\omega<1$, implying the proton radius diminishes for $\omega_0<\omega\rightarrow 1$, i.e. for $T\rightarrow T_c$. The proton radius ceases to diminish by some blocking mechanism as, for example, the Pauli blocking. In the TE, this effect is observed in the vanishing of the correlation term. 

The positive energy, on the other hand, is dominant for $1<\omega$. In this case, the proton radius starts to increase for $T_c<T$. Then, the correlation term in the TE becomes relevant to explain the proton thermal evolution. When the Hagedorn temperature is achieved the process should stop since the proton dissociates into a quark soup. 

An analogy between the thermal evolution of our model and the total cross section behavior according to the collision energy is performed. The negative energy in the von Laue result is associated with the odderon exchange and the positive one with the pomeron exchange. The hollowness effect can also be understood in terms of the von Laue result. A detailed study near the transition temperature can be performed by using the approach proposed in \cite{T.S.Biro.A.Jakovac.Phys.Rev.Lett.94.132302.2005,G.Wilk.Z.Wlodarczyk.Eur.Phys.J.A.48.161.2012,G.B.Bagci.T.Oikonomou.Phys.Rev.E.88.042126.2013,T.S.Biro.G.G.Barnafoldi.P.Van.Physica.A.417.215.2015,T.S.Biro.Physica.A.392.3132.2013} and will be developed elsewhere.



\section*{Acknowledgments}

SDC thanks to UFSCar by the financial support and AMA thanks to IFSC-USP.

\end{document}